\documentclass[preprint]{aastex}
\usepackage{emulateapj5}
\newcommand{\keV}{\rm keV}
\newcommand{\kpc}{\rm kpc}
\newcommand{\msun}{M_{\odot}}
\newcommand{\mdot}{\dot M}
\newcommand{\rtr}{R_{\rm tr}}
\newcommand{\rout}{R_{\rm out}}
\newcommand{\rs}{R_{\rm Schw}}
\newcommand{\nt}{XTE~J1118$+$480}
\submitted{To appear in the Astrophysical Journal}

\shorttitle{Modeling the Low State Spectrum of the X-Ray Nova XTE~J1118+480}
\shortauthors{Esin et al.}

\eqsecnum

\begin{document}

\title{Modeling the Low State Spectrum of the X-Ray Nova XTE~J1118+480}
\author{Ann A. Esin\altaffilmark{1}\altaffilmark{,2}, 
Jeffrey E. McClintock\altaffilmark{3},}
\author{Jeremy J. Drake\altaffilmark{3}, Michael R. Garcia\altaffilmark{3}, 
Carole A. Haswell\altaffilmark{4},}
\author{Robert I. Hynes\altaffilmark{5}, Michael P. Muno\altaffilmark{6}}
\altaffiltext{1}{Caltech 130-33, Pasadena, CA 91125; aidle@tapir.caltech.edu}
\altaffiltext{2}{Chandra Fellow}
\altaffiltext{3}{Harvard-Smithsonian Center for Astrophysics, Cambridge, MA 
02138}
\altaffiltext{4}{Department of Physics and Astronomy, Open University, 
Walton Hall, Milton Keynes, MK6 7AA, UK}
\altaffiltext{5}{Department of Physics and Astronomy, University of 
Southampton, Southampton, SO17 1BJ, UK}
\altaffiltext{6}{Center for Space Research, MIT, Cambridge, MA 02139}

\begin{abstract} 
Based on recent multiwavelength observations of the new X-ray nova
\nt, we can place strong constraints on the geometry of the accretion
flow in which a low/hard state spectrum, characteristic of an
accreting black hole binary, is produced.  We argue that the absence
of any soft blackbody-like component in the X-ray band implies the
existence of an extended hot optically-thin region, with the
optically-thick cool disk truncated at some radius $R_{\rm tr} \ga 55
\rs$.  We show that such a model can indeed reproduce the main
features of the observed spectrum: the relatively high optical to
X-ray ratio, the sharp downturn in the far UV band and the hard X-ray
spectrum.  The absence of the disk blackbody component also
underscores the requirement that the seed photons for thermal
Comptonization be produced locally in the hot flow, e.g. via
synchrotron radiation.  We attribute the observed spectral break at
$\sim 2\,\keV$ to absorption in a warm, partially ionized gas.

\end{abstract} 


\section{Introduction}
\label{intro}
There is now solid evidence for the presence of black holes within a
number of X-ray binaries.  A low/hard state, characterized by a
power-law spectrum with photon index of $1.4-1.8$ and extending to
$\sim 100\,\keV$, is observed both in transient systems (X-ray novae,
e.g. GRS~1124-68, GRO~J0422+32) and persistent systems (e.g. Cyg~X-1,
LMC~X-3).  It is generally assumed that these spectra are produced by
thermal Comptonization of seed photons in the vicinity of an accreting
black hole.  However, the geometry of the accretion flow and the
source of soft photons currently remains a point of debate.  Recently
two competing pictures have emerged.  In the accretion disk corona
model, a cool thin disk is embedded in a hot corona (thought to be
heated by magnetic flares), which produces the hard power-law emission
via inverse Compton scattering of the disk photons \citep*[e.g.][and
references therein]{grv79, ham91, hmg94, pos96}.  In the other
scenario, the thin disk is truncated at some large inner radius and
the bulk of the emission forms in a hot quasi-spherical accretion
flow through Comptonization of locally produced synchrotron and
bremsstrahlung radiation \citep*[e.g.][]{sle76, ich77, emn97, dea97,
enc98}.

The simplest way to distinguish between these two models would be to
observe directly the emission from the inner edge of the thin disk.
However, at accretion rates for which the low state is generally
observed, $\la10\%$ of Eddington \citep[see e.g.][]{now95}, the thin
disk emission peaks at energies $\la 0.5\,\keV$.  This region of the
spectrum is strongly affected by interstellar absorption and, until
recently, has been at the edge of the sensitivity range for existing
X-ray detectors.  In the past, the attempts to distinguish between
these two scenarios have been based mostly on modeling the X-ray
reflection features (e.g. \citealp*{gea97, zds97, zds98, doz99},
though see also \citealp*{dea97}), the so called ``reflection bump''
centered at $\sim 30\,\keV$, and the iron fluorescence line at
$6.4\,\keV$ \citep{gur88, gef91}.  However, the interpretation of
these features is complicated by uncertainties in the models
\citep{rfy99, nkk00}.

Currently, this state of affairs is changing.  Both Chandra and XMM
are capable of observing soft X-rays down to $\sim 0.2\,\keV$ with
good energy resolution.  In addition, a new black hole X-ray nova --
\nt\ -- was discovered on 2000 March 29 \citep{rea00} in a region with
exceptionally low interstellar absorption \citep{gea00}, which made
possible the first extreme UV observation of an X-ray nova
\citep*[hereafter HEA]{hmh00}.  Based on near-simultaneous optical,
UV, and X-ray observations, \citeauthor{hmh00} concluded that the
source appeared to be in a low state (typical of black hole systems)
with a power-law spectrum extending down to $\sim 100\,{\rm eV}$; the
photon index was $1.8\pm0.1$.  The authors suggested that these data
are inconsistent with the thin disk extending to the last stable
orbit.

In this paper we present detailed modeling of the combined
near-simultaneous HST, EUVE, Chandra and RXTE observations of \nt\
\citep*[hereafter PI]{mea01a}.  The addition of the Chandra LETG
spectrum bridges the gap between the EUV and X-ray data discussed by
\citeauthor{hmh00}, providing us with a complete description of this
previously inaccessible part of the spectrum.  This is especially
important because of the claim based on a 2000 May 11-12 ASCA
observation \citep*{yud00} that the spectrum of \nt\ showed a slight
soft excess that is consistent with a blackbody component of
temperature $0.2\,\keV$.  Clearly this is inconsistent with the
Chandra data.  Moreover, in view of the source stability in radio,
optical and X-rays (\citeauthor{mea01a}), the spectrum is not
likely to have changed between the two observations.  We argue that
the combined spectrum allows us, for the first time, to place direct
limits on the low state accretion geometry in the vicinity of the
central object.

In \S\ref{results} we show that a model based on an
advection-dominated accretion flow surrounded by a truncated thin disk
(described in \S\ref{model}) gives an excellent fit to the combined
optical, UV and X-ray data.  We argue in \S\ref{disc} that the
observed spectrum rules out the presence of the standard thin
accretion disk within $\sim55$ Schwarzschild radii ($\rs$) and thereby
presents a fundamental problem for the accretion disk corona models.

\section{Model Overview}
\label{model}

The basic model we use is described in detail in \citet{emn97} and
\citet{enc98} and is briefly summarized below.  Here we have added
fully relativistic flow dynamics \citep{gap98}, gravitational
redshift \citep{nea98} and the kinematic red- blue-shift of the
emission due to the relativistic motion of the gas near the black hole
horizon.  In this model, the material transfered from a mass-losing
secondary initially forms an optically thick, cool disk \citep{shs73}
outside some radius, $\rtr \ga 30-100 \rs$.  Inside this radius the
gas accretes via an optically thin, hot advection-dominated accretion
flow (ADAF; see \citealt*{nmq98} and references therein).  

The emission from such an accretion flow consists of two components.
The first of these is the standard multicolor blackbody spectrum from
an optically thick and cool outer disk.  The effective temperature as
a function of radius is determined by the viscous dissipation within
the disk and irradiation of the disk surface by the inner ADAF.  To
compute the irradiation heating rate we assume that the thickness of
the disk, $H$, is equal to the pressure scale height.  The temperature and
normalization of the thin disk spectrum is then determined mainly by
four parameters: the black hole mass $M$, the inner and outer radii of the
disk, $\rtr$ and $\rout$, and $\mdot_{\rm disk} \sim \mdot$.

The second emission component is produced in an optically thin, hot
ADAF, where the electrons have temperatures of order
$10^9-10^{10}\,{\rm K}$.  In the presence of strong magnetic fields,
such a gas cools through inverse Compton scattering of thermal
synchrotron radiation.  The spectrum is roughly a power-law with a
high energy thermal cutoff at $100-200\,\keV$ and a low energy cutoff
in the IR/optical band due to synchrotron self-absorption.  The slope
of the power-law spectrum is determined mainly by $\mdot/(\alpha^2
\mdot_{\rm Edd})$, where $\alpha$ is the standard viscosity parameter,
which we take to be $\alpha=0.25$.  At $\mdot \sim 0.06 \mdot_{\rm
Edd}$, the photon index is close to $1.4$ and it increases with
decreasing $\mdot$.  The high energy cutoff is determined by the
electron temperature in the inner part of the flow, which depends most
sensitively on $\beta$, the ratio of the gas pressure to the total
pressure (from gas and magnetic fields).  
 
\subsection{Constraints on Model Parameters}
\label{binary}
\setcounter{footnote}{0}

\nt\ entered quiescence in October 2000, and subsequent spectroscopy
of the optical counterpart placed fairly strong constraints on $M$,
the distance to the system, $d$, and its binary inclination, $i$.
\citet{mea01b} and \citet{wea01} independently confirmed the orbital
period to be $4.1\,{\rm hours}$ (as previously suggested by
\citealt{cea00,pat00,uea00,dea01}) and determined the value of the
mass function to be $f(M) \simeq 6.0\msun$.  \citet{mea01b}
constrained the spectral type of the companion to lie in the range
K5V-M1V, implying that its mass is a fraction of solar, and estimated
the distance to be $d=1.8\pm 0.6\,\kpc$.  They also argued that the
observed photometric modulation, if attributed to ellipsoidal shape of
the companion, implies $i>40^{\circ}$.

In our model, for fixed values of $\alpha$ and $\beta$, the value of
the mass accretion rate in Eddington units is set by the observed
spectral slope.  The relative normalization of the optical and X-ray
emission components depends only on $M$ and $i$\footnote{The thin disk
emission, which dominates in the optical, varies with inclination as
$\cos{i}$, while we assume here that the ADAF geometry is close to
spherical and so X-ray emission has practically no $i$-dependence.}.
Because the mass function $f(M)$ provides another relation between these two
parameters, the best fit to the data uniquely determines both $M$ and
$i$, once $\mdot$, $\alpha$ and $\beta$ are fixed ($f(M)$ also
has a weak dependence on the mass of the companion, which we
set to $0.4 \msun$).  In our calculations we treat $d$ and $\rtr$
as free parameters; the former controls the overall normalization of
the spectrum, and the latter determines the position of the thermal
emission peak.

Finally, the outer radius of the accretion disk, $\rout$, is estimated
using \citeauthor{pac71}'s \citeyearpar{pac71} formula:
$\rout = 3\times 10^4 \rs ({10\msun}/{M})^{2/3}$,
where we have taken the binary mass ratio to be $<0.1$
\citep{dea01,mea01b,wea01}, and assumed that the accretion disk fills
$80\%$ of the primary's Roche lobe.

\section{Results}
\label{results}
Using the model described in \S\ref{model}, we have computed the best
fit spectrum for the canonical parameter values $\alpha=0.25$ and
$\beta=0.5$.  This spectrum is shown as a dashed line in
Fig. \ref{fig1} together with the data from \citeauthor{mea01a}.  With
$\mdot = 0.035 \mdot_{\rm Edd}$ and $\rtr = 65 \rs$ we can easily
reproduce both the photon index of 1.8 required by the data in the
range $2-30\,\keV$ and the steep slope in the EUV band.  To reproduce
the ratio of X-ray and optical fluxes with this model, we need a very
massive black hole, $M=17\msun$, with a disk inclined at
$i=45^{\circ}$, at a distance $d=2.9\,\kpc$.

The UV-band energy slope of the model spectrum is fairly close to the
canonical thin disk value of $1/3$.  In fact, it
is slightly flatter, as required by observations (\citeauthor{hmh00}),
due mainly to a contribution from the self-absorbed synchrotron
emission below $1.5\times 10^{15}\,{\rm Hz}$.  However, the disk
emission steepens considerably below $\sim 5\times 10^{14}\,{\rm Hz}$,
so that the red end of the HST spectrum and the IR data points clearly
lie above our model spectrum.  This discrepancy decreases somewhat if
$\rout$ is a factor of 3 greater than the value we derived in
\S\ref{binary}.  However, the presence of strong non-thermal radio
emission in this system\footnote{Not shown in Fig. \ref{fig1}, but
see \citeauthor{hmh00} and \citeauthor{mea01a}} suggests that the 
extra contribution to the IR (and possibly optical) emission comes
from an entirely different source, perhaps an accretion disk outflow
or a jet \citep[e.g.][]{fea00}, or a small population of non-thermal 
electrons in the ADAF \citep[e.g.][]{opn00}.

More important discrepancies between this model and the data are (1)
the fact that our theory predicts a turnover in the hard X-ray band
with a characteristic $e$-folding energy $\sim 180\,\keV$, while the
summed RXTE data clearly show a power-law spectrum continuing to much
higher energies; and (2) a lack of a spectral break at $\sim
2\,\keV$ in the model spectrum.

\includegraphics{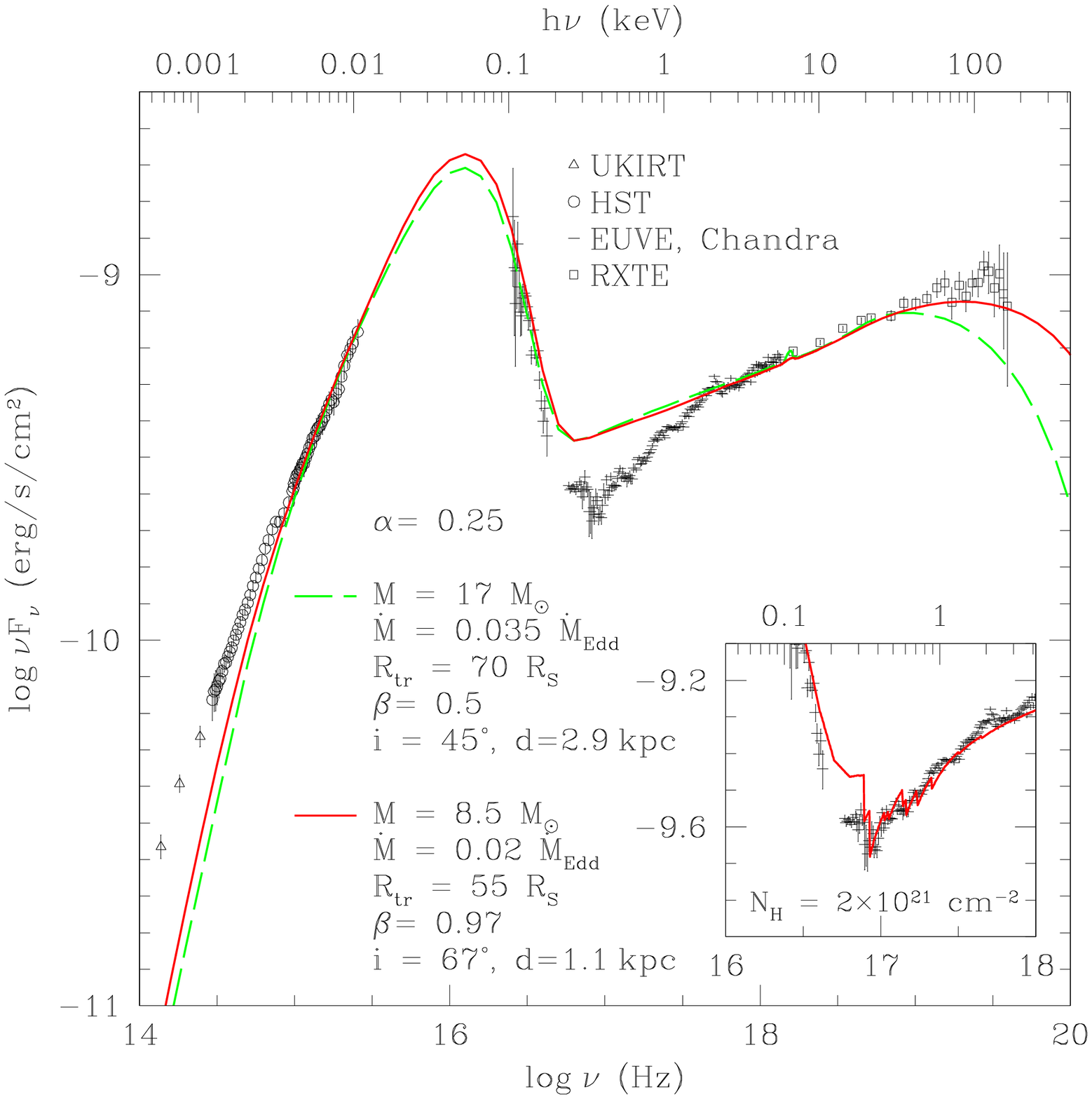} 
\vskip 3.2in 
\figcaption{\label{fig1}The model spectra for different parameter
values are shown together with the data from \citeauthor{mea01a}, corrected
for interstellar absorption with $N_H = 1.15\times 10^{20}\,{\rm cm^{-2}}$
(see \S\ref{disc}).  
The model in the inset incorporates the effects of a warm absorber
(see text for details).}

A spectrum calculated assuming a higher value of $\beta=0.97$ (shown
by the solid curve on Fig. \ref{fig1}) gives a much better fit to the
data in the hard X-ray band.  Higher $\beta$ implies lower magnetic
field strength in the accreting gas, and thus less synchrotron
emission.  The energy balance in the flow then requires a higher
electron temperature to keep up the cooling rate, which moves the
spectral cutoff to $\sim 350\,\keV$. In addition, increasing the value
of $\beta$ causes a decrease in the viscous dissipation rate in the
ADAF\footnote{When $\beta \rightarrow 1$, the ADAF adiabatic index
$\gamma \rightarrow 5/3$, which corresponds to a non-rotating
accretion flow with no dissipation \citep{nay94,esi97}.}, and thus a
decrease in its overall luminosity.  As a result, this model can
reproduce the observed X-ray to optical flux ratio with a relatively
high inclination of $i=67^{\circ}$ and smaller black hole mass,
$M=8.5\msun$, and distance, $d=1.1\,\kpc$.  Because the accretion flow
is hotter than with $\beta=0.5$, the mass accretion rate required to
reproduce the observed slope in the X-ray band decreases to $\mdot =
0.02\mdot_{\rm Edd}$.  The transition radius in this model is $\rtr =
55\rs$.

The low $B$-field model leaves unexplained the broad dip in the data
centered around $0.3\,\keV$.  In general, since X-ray emission in our
model is produced in an optically thin region with a nearly constant
electron temperature, the break in the power-law spectrum is difficult
to reproduce; moreover, any scenario that relies on thermal
Comptonization of low-energy seed photons to produce X-ray emission
will encounter the same difficulty.  Here we suggest that this feature
in the observed spectrum is due to metal absorption in a partially
ionized gas with $T\sim {\rm few} \times 10^5\,{\rm K}$.  At this
temperature H and He are completely ionized, and since these elements
are the main contributors to opacity below $200\,{\rm eV}$, the
presence of extra material does not affect the EUVE data.  The
abundant metals, C, N, O, Ne, etc., on the other hand, will be only
partially ionized and their $K$-edge energies fall precisely in the
range where absorption seems to be the strongest.  Such a warm
absorber model is commonly proposed to explain similar features in AGN
spectra \citep[see][and references therein]{kom00}.

To illustrate this suggestion, we have approximated the effect of
$K$-edge absorption by C, N, O, Ne and Mg, together with $L$-shell
absorption by Fe, on our high-$\beta$ model described above.  Because
the locations of the strongest absorption features near $300\,{\rm eV}$
are close to the $K$-edge energies of CIII and CIV, we have included
only the contribution from these stages of ionization for all the
metals. (We realize that this is a major approximation and that a
realistic model would require solving the full ionization balance
equations).  The result is shown in the inset in Fig. \ref{fig1} for
$N_H \sim 2\times 10^{21}\,{\rm cm^{-2}}$. (All elements, except O,
are assumed to have solar abundances; because of the absence of a
strong O edge, we have reduced its abundance by a factor of 3.)
Because of its simplicity, this is not meant to be a real model;
nevertheless, it demonstrates that photoelectric absorption (with its
characteristic frequency dependence of optical depth, $\tau\propto
\nu^{-8/3}$) can indeed reproduce the general shape of the soft X-ray
spectrum.  Note that the discrepancy between our toy model and
the data at energies below the C $K$-edge is most likely due to the
contribution of $L$-shell absorption, which we neglected here.

It is important to point out that the thin disk is {\em not} a
significant source of seed photons for inverse Compton cooling of the
gas in the inner ADAF.  The synchrotron emission dominates as a soft
photon source due to the large inner disk radius, even with
$\beta=0.97$.  In fact, less than $1\%$ of the thin disk photons
penetrate the ADAF within $30\rs$, the region where most of the
cooling occurs.  The small angular size of the disk, as seen by the
inner ADAF, also accounts for the weakness of the reflection features
in the model spectrum.  The EW of the Fe K$_{\alpha}$ line calculated
assuming a neutral cold outer disk is $\sim 14\,{\rm eV}$, consistent
with the 2-$\sigma$ upper limit of $22\,{\rm eV}$ from the Chandra
data.

\section{Discussion}
\label{disc}

The steep slope of the observed spectrum in the EUV band and the
smooth power-law emission at higher energies strongly suggest that
near $0.1\,\keV$ we are seeing the Wien tail of the blackbody emission
from the inner disk edge.  \citeauthor{mea01a} estimated the
temperature of this component to be around $20-25\,{\rm eV}$.  If we
assume that the accretion proceeds through a thin disk extending to $3
\rs$, this inner edge temperature implies a disk accretion rate of
$\mdot \sim 2 \times 10^{-7} (M/10 \msun) \mdot_{\rm Edd}$.  Even at $10\%$
efficiency the bolometric luminosity from an accretion disk with such
low $\mdot$ is of order $3\times 10^{32} (M/10 \msun)^2\,{\rm erg\ s^{-1}}$,
nearly four orders of magnitude lower than the inferred luminosity of
\nt, $L\ga 7\times 10^{35} (d/1.1\,\kpc)^2\,{\rm erg\ s^{-1}}$.
This result is practically unchanged even if the bulk of accretion
occurs in the corona above the disk, since as long as the disk is
optically thick it will be heated by irradiation
\citep*[e.g.][]{dwb97}.  For example, the magnetic flare model
proposed by \citet*{mdf00} predicts the peak of the disk emission near
$0.2\,\keV$, in clear disagreement with both the EUVE and Chandra
data.  At higher accretion rates, the emission from the inner disk
could be hidden by an optically thick corona or occulted by an outer
rim of the disk itself. However, the first explanation is ruled out if
we assume that the same corona produces power-law emission via inverse
Compton scattering, since the cutoff temperature $\ga 250\,\keV$ and
photon index 1.8 imply an optical depth in the corona of $\la 0.4$.
On the other hand, occultation of the inner part by the disk rim is
possible only if the system is nearly edge-on with $90^{\circ} - i =
5.7^{\circ} H_{\rm out}/0.1 \rout$.  Such a fine-tuned orientation
seems unlikely.  Moreover, the lack of observed X-ray eclipses due to
the companion places an upper limit of $80^{\circ}$ on the binary
inclination.  Thus, occultation by the disk rim can play a role only
if the outer disk is very thick\footnote{Note that the lack of
eclipses indicate that it is not sufficient to have a warped outer
disk; instead $H_{\rm out}$ must reflect a true thickness of the
disk.}, with $H_{\rm out}/\rout \ga 0.2$.  Overall, we feel that a
thin disk truncated near $\sim 50-70\rs$ provides a much more
plausible explanation of the observed data.

The interstellar absorption column adopted in this paper,
$N_H=1.15\times 10^{20}\,{\rm cm^{-2}}$, is $10\%$ smaller than the
preferred value of $1.3\times 10^{20}\,{\rm cm^{-2}}$ quoted in
\citeauthor{mea01a}.  The smaller $N_H$ gives a better fit to our
theoretical spectrum in the EUV energy band, since the temperature
profile of the inner disk is shallower in our model than in the
multicolor disk blackbody model used by \citeauthor{mea01a}.  We also
cannot exclude a possibility that the absorption column is even
smaller, e.g. $N_H = 0.75\times 10^{20}\,{\rm cm^{-2}}$ (as suggested
by \citeauthor{hmh00}).  In this case, the EUVE spectrum will form a
continuation of the power-law component, forcing the transition radius
out to $\rtr \ga 160 \rs$.  Thus, lower $N_H$ would only strengthen
our overall conclusions.

\citet{dmp99} have suggested that low-frequency QPO's seen in many
black hole systems give an upper limit on the inner radius of a thin
accretion disk.  A QPO at a frequency $0.07-0.15\,{\rm Hz}$ was seen
in optical, UV and X-ray monitoring of \nt\ \citep*{pat00,
rsb00,hsp00,yud00,wea00}.  The formalism of \citeauthor{dmp99} would
then predict a transition radius at $\le 13-15\, (6\msun/M)^{2/3}
\rs$, which is clearly inconsistent with our result in
\S\ref{results}, i.e.  $\rtr \ga 55 \rs$, for any allowed
value of the black hole mass in \nt.  It is also unlikely that
this discrepancy is due to source variability, since the QPO was
observed both before and after the observations discussed here, and
the source emission was quite stable in both the optical and X-ray
bands for several weeks around this time (\citeauthor{mea01a}).

A somewhat peculiar feature in the X-ray spectrum of \nt\ is the
apparent absence of a turnover at high energies in the summed HEXTE
data taken between April 13 and May 15 (see Fig. \ref{fig1}).  By
comparison, the $e$-folding energy observed during the 1992 outburst
of GRO~J0422$+$32 (a black hole X-ray nova with a similar,
persistently hard X-ray spectrum) spanned the range $107-130\,\keV$
\citep{enc98}.  The difference in spectra therefore suggests a weaker
magnetic field (i.e. larger $\beta$) in XTE~J1118$+$480 than
previously used to model GRO~J0422$+$32.  This is somewhat disturbing,
since $\beta$ is determined by the basic physics of an accretion flow
and is not expected to vary significantly from system to system (at
least at similar mass accretion rates). Thus, a confirmation of the
HEXTE observations is very important.  Preliminary data analysis of
Beppo-SAX observations of XTE~J1118$+$480 on 2000 April 14 suggests
that a high energy cutoff is indeed present below 300~keV
\citep{fro00}.  If the exact value of the $e$-folding energy is
determined, it will be crucial for constraining the electron
temperature and thus, the value of $\beta$.  On the other hand, if it
is convincingly shown that the high energy emission continues above
300~keV without a break, our model (and in fact any thermal emission
model) can be effectively ruled out.

The combination of known mass function and simultaneous optical and
X-ray observations allowed us to constrain the black hole mass, binary
inclination and distance to \nt\ in the context of our model.  For
$\beta = 0.5$ we find $d=2.9\,\kpc$, considerably larger than the
estimate of \citet{mea01b} and \citet{wea01}; for $\beta=0.97$ our
value, $d=1.1\,\kpc$, is in better agreement with observations in
quiescence and with the distance estimate based on optical
spectroscopy in outburst \citep{dea01}.  If $i$, $M$, and $d$ are
better constrained by future observations, their values can be used to
constrain $\beta$ and $\alpha$ in the ADAF model.

Our best-fitting model spectrum with $\beta=0.97$ corresponds to
optical magnitudes $V=13.1$ and $R=13.0$ (estimated using standard
photometric system conversion from \citealt{all73,wam81}),
respectively $\sim 0.2$ and $\sim 0.4$ magnitudes dimmer than what is
observed.  The rest of the optical emission probably represents a
contribution from some non-thermal source (see discussion in
\S\ref{results}).  Our magnitude estimates are of course upper limits,
since we normalized our model in such a way so as to reproduce all of
the observed UV flux.  It is possible that the non-thermal emission
component makes a significant contribution to the HST spectrum even at
highest frequencies.  In this case the disk emission is dimmer than we
estimated here and the best fit black hole mass is somewhat smaller.
For example, if the thin disk contributes only $60\%$ of the flux at
$2\times 10^{15}\,{\rm Hz}$, our $\beta=0.97$ model requires
$M\simeq7.3\msun$, $i\simeq77^{\circ}$, and $d\simeq1.0\,\kpc$.
 
It is tempting to ascribe the spectral break below $2\,\keV$ to
absorption by an outflowing gas, which is also responsible for the
radio and IR excess emission.  For a spherically-expanding accretion
disk wind, the H column density estimated in \S\ref{results} implies a
density $n (R_{\rm w}) \sim 3 \times 10^{10} (\rout/R_{\rm w})\,{\rm
cm^{-3}}$.  This is consistent with a few percent of the mass
transfered from the companion being launched as a wind from radius
$R_{\rm w}$ at a fraction of a Keplerian velocity.  However, the bulk
of metal atoms in such a wind will be fully photoionized by the
UV/X-ray photons.  To avoid this problem, the absorbing material has
to be located further away from the system.  Since the total mass
requirement increases as $\propto N_H R^2$, in this case the observed
column density may be due to the accumulation of material ejected
during many past outbursts.  From a simple photo-ionization balance
calculation, we estimate that CIV will be the main ionization state of
C outside $\sim 2\times 10^{15}\,{\rm cm}$.  For a spherical outflow,
this implies a lower limit of $10^{-3} (T/10^5\,{\rm K}) \msun$ on the
total mass of the absorbing material at temperature $T$.  For a
time-average mass transfer rate of $\mdot_{\rm T} \sim 10^{16}\,{\rm
g\ s^{-1}}$, expected in \nt\ from binary evolution theory
\citep{kkb96}, this amount of material can easily be lost in a wind
over $\sim 10^7\,{\rm yr}$ (a much shorter time than the typical age
of a low mass X-ray binary).

The estimates above depend strongly on the assumed metal abundances.
Based on the optical/UV spectroscopy of \nt\ in outburst,
\citet*{hhk01} and \citet{dea01} suggest that the accreting material
has low abundances of C and O as compared to N; they attribute this to
CNO processing.  This interpretation may help to explain the lack of a
prominent O edge in the X-ray spectrum of \nt\ but it is not consistent
with our interpretation of the strong $\sim 300\,{\rm eV}$ feature as
a C $K$-edge.  

Finally, it is also possible that the absorbing material is not
associated with the system.  In the future, as our understanding of
the detector response continues to improve, we plan to do a more
detailed analysis of the Chandra spectrum, placing limits on the
position and strength of individual absorption edges.  Together with a
detailed model of the absorbing material, which would include a
realistic calculation of the ionization balance, this work should shed
some light on the physical properties and the origin of the warm
absorber.
 
\section{Summary}
\label{summary}
The ADAF model proposed by \citet{emn97} to explain the low state
spectra of accreting black holes can explain the observed spectrum of
the new transient source, \nt, in outburst.  In the case of Nova
Muscae, due to the lack of UV and soft X-ray data, \citet{emn97} could
not constrain the position of the transition region between an outer
thin disk and an inner hot ADAF, beyond requiring that it must lie
outside $\sim 30\rs$ from the black hole \citep{enc98}.  The
unprecedented multiwavelength observations of \nt\
(\citeauthor{hmh00}; \citeauthor{mea01a}) allow a direct determination
of the position of the inner edge of the thin disk.  We show that the
simultaneous optical, UV and X-ray data can be explained by an
ADAF+disk model with a transition radius at $55 \rs$, accreting onto a
$\sim 9 \msun$ black hole at a rate $2\%$ of Eddington, and a source
distance of $d \sim 1.1\,\kpc$.  We find that we
need a fairly high value of $\beta$ in our model, i.e. weak magnetic
fields, to reproduce the unusually high cutoff energy found in the
HEXTE observations of \nt.  Finally, we attribute the hardening of the
soft X-ray spectrum below $2\,\keV$ to the presence of a warm
absorbing medium between us and the system.

\acknowledgements AAE thanks Jonathan Kawamura and Roger Blandford for
helpful suggestions and discussions, and Kristen Menou for help with
high-$\beta$ dynamical models.  AAE was supported by NASA through
Chandra Postdoctoral Fellowship grant \#PF8-10002 awarded by the
Chandra X-Ray Center, which is operated by the SAO for NASA under
contract NAS8-39073. JEM, MRG and JJD acknowledge NASA support from
grant DD0-1003X and contract NAS8-39073.  CAH \& RIH are supported by
grant F/00-180/A from the Leverhulme Trust.

\bibliography{bhn}

\end{document}